
\magnification=\magstep1
\baselineskip = 18 true pt
\hsize =  17 true  cm
\vsize =  23 true cm
\centerline {{\bf A Simple Lattice Version of the Nonlinear Schrodinger
Equation and its Deformation}}\centerline {{\bf with Exact Quantum Solution
 }}
\vskip 1 true cm
\centerline {A Kundu  }
\centerline {{\it Physikalisches Institut Der Universit\"at Bonn}
}\centerline{{\it Nussallee 12, 53115 Bonn, Germany}}
\centerline
{{\it Permanent address:
Theoretical Nuclear Physics  Division,}}\centerline {{\it
  Saha
Institute of Nuclear Physics }}
\centerline {{\it
 Sector 1, Block-AF, Bidhannagar, Calcutta-700 064, India }}
\centerline { O.Ragnisco}  {\it
\centerline { Universit\`a di Roma III, Dipartimento di Fisica }
\centerline {{\it and}}
\centerline {{\it Istituto Nazionale di Fisica Nucleare, Sezione di Roma,
P.le A.Moro 2, 00185 Roma,Italia} }}
                        \vskip 1 true cm

\vskip 2 true cm
\noindent {\it Running title : }
Simple lattice NLSE \hfil\break
\noindent {\it Subject Classification : }
Mathematical Physics \hfil\break
\vskip 1 true cm
\noindent {\bf Abstract}
A   lattice version of quantum nonlinear Schrodinger
(NLS) equation  is considered, which has  significantly simple form
and fullfils most of
 the  criteria desirable for such  lattice variants
of field models. Unlike most of the known lattice NLS,
the present model belongs to a class
 which  does not exhibit the usual symmetry properties. However
this lack of symmetry itself seems to be responsible for
the remarkable simplification of the
relevant objects in the theory, such as the Lax operator, the
 Hamiltonian and other commuting conserved quantities as well as
their  spectrum.
 The
model allows exact quantum solution
through algebraic Bethe ansatz and also a straightforward
 and natural generalisation to the
vector case, giving thus a new  exact lattice version
of the vector NLS model. A deformation  representing a new
quantum integrable system involving
Tamm-Dancoff like $q$-boson
 operators is constructed.
\vfil\eject

\noindent {\bf 1. Introduction}

 Among integrable systems discrete models represent a special class,
 interest to which has been revived in recent years [1-3]. In the
 context of quantum integrable systems, apart from their own right as
 solvable quantum lattice models, they also play an important role by providing
 lattice regularised versions of the corresponding continuum models. Thus
 the lattice nonlinear Schrodinger (NLS)  [4-9], lattice  sine-Gordon [7]
 models etc. are useful for finding out
 the exact quantum solutions of the related
 field models through the quantum inverse scattering method
  (QISM) [10]. Moreover, these lattice
 versions are  often able to unveil the hidden algebraic
 structures of the original field models [1,11].

 Ideally a candidate for such quantum
 integrable discrete models
 represented by the Lax operator $L(n\mid \lambda)$
 should fulfill the following basic criteria:

 {\it  \quad i) It should satisfy exactly the
 quantum Yang-Baxter equation
(QYBE)
$$ R(\lambda - \mu)~ L(\lambda)\otimes L(\mu ) = L(\mu )\otimes L(\lambda)~
R(\lambda
- \mu), \eqno(1.1)  $$

 \quad ii) It should have  the same quantum $R$-matrix as the corresponding
 field model.

\quad iii) The discrete Lax operator should yield the continuum one
  at $\Delta \rightarrow 0:$

 \quad $L(n\mid \lambda) \rightarrow {\bf 1}+
  \Delta {\cal L}(x,\lambda)$, $\Delta$ being the lattice constant.

  \quad iv) The Hamiltonian of the discrete model should be `local'
  and give back the local

\quad field model at the continuum limit.}

Moreover, as a desirable physical requirement
{\it the Lax operator as well as the Hamiltonian
  and the energy spectrum should be as simple
  as possible}.

  A discrete version of NLS model was first suggested by Ablowitz
  and Ladik [5]. However the corresponding $R$-matrix , a key object in
  QISM,  is expressed through
  trigonometric functions [12] and does not coincide with the well known
  rational $R$-matrix  of the NLS field model, which reads:
$$
  R (\lambda) =  \lambda {\bf P} -i\kappa  {\bf I} =
\pmatrix {
 { \lambda - i\kappa
  } & { } & { } & { } \cr { } &
{- i\kappa }& {\lambda }& { } & { }
\cr{ }& { \lambda } &
{- i\kappa
 }& { } & { }\cr { } & { } & { }
&{ \lambda - i\kappa
  } \cr }
.\eqno (1.2)$$
That is, criterion ii) laid down above  is not satisfied.
Subsequently, a different discrete NLS was proposed [6],
   constructed through the Holstein-Primakoff transformation
   (HPT) applied to an
   infinite dimensional irreducible representation of $su(2)$
   with the classical Lax operator given by
 $$ L(n \mid\lambda) =
 \pmatrix {
 {c - {i\over 2 } \lambda \Delta +{\kappa \over 2 } \Delta^2 \psi(n)^*
 \psi(n) }
&{ -i \psi(n)^*
\Delta  \sqrt{\kappa  (c+ {\kappa \over 4 }
\Delta^2 \psi(n)^*\psi(n))  }   }
\cr
{ i  \Delta \sqrt{\kappa  (c+ {\kappa \over 4 }
\Delta^2 \psi(n)^*\psi(n))  } \psi(n)  }
&{ c + {i\over 2 } \lambda \Delta +{\kappa \over 2 } \Delta^2 \psi(n)^*
 \psi(n) } \cr }
 ~ , \eqno (1.3) $$ related to spin parameter $~~~s=-{2 \over {\kappa \Delta}}~
{}~$
 with $c=1$ and
 canonical Poisson  brackets

\noindent
 $~~~~\{ \psi(n),\psi(m)^* \} = \delta_{nm}~~~ $.

 This model is free from the earlier drawback, namely it satisfies ii),
  though  it now fails to fulfill the locality criterion iv) at the
  quantum level. As a remedy another version of lattice NLS was introduced
  [7], represented by a Lax operator which explicitly depends on
  lattice points and may be expressed as
  a product $l(n\mid \lambda)=L(2n\mid \lambda)
  L(2n-1\mid \lambda)$, where $L(n\mid \lambda)$ is taken as (1.3) with
  $c=1+{1\over 4}(-1)^n\kappa\Delta$, that is:

  $L(n\mid \lambda)=$\hfil\break
 $$\pmatrix {
 { 1+{1\over 4}(-1)^n\kappa\Delta
  - {i\over 2 } \lambda \Delta +{\kappa \over 2 } \Delta^2 \psi(n)^*
 \psi(n) }
&{ -i \psi(n)^* \Delta \sqrt{\kappa   (1+{1\over 4}(-1)^n\kappa\Delta
 + {\kappa \over 4 }
\Delta^2 \psi(n)^*\psi(n))  }   }
\cr
{ i  \Delta \sqrt{\kappa  (1+{1\over 4}(-1)^n\kappa\Delta
 + {\kappa \over 4 }
\Delta^2 \psi(n)^*\psi(n))  } \psi(n)  }
&{  1+{1\over 4}(-1)^n\kappa\Delta
 + {i\over 2 } \lambda \Delta +{\kappa \over 2 } \Delta^2 \psi(n)^*
 \psi(n) } \cr } .\eqno (1.4)
    $$

 This model satisfies  the required criteria,
but  looks involved to comply with the
physical requirement  of  simplicity. Finally, in a further investigation [8],
  a relatively simpler model was
  proposed, where the Lax operator was given directly by
  (1.4). However  the important criterion iii) is now not fulfilled
 and the simplification
achieved is also not  fully satisfactory, as it is evident from  the form of
the Lax operator and structure of the following Hamiltonian:
$$
H=-{4\over3\kappa\Delta^3} \sum_n^N(t_n+t^\dagger_n+{8-\kappa\Delta\over
8-2\kappa\Delta})+({4\over3\Delta^2}+{\kappa^2\over12})\sum_n^N  \Delta
\psi^\dagger_n\psi_n ,\eqno (1.5)$$
where the local density $t_n$ again has different expressions, depending
whether it corresponds to even or odd sites.  For
odd $n$ it takes the form
$$\eqalign {t_n
=(\alpha^\dagger(n+2)\alpha(n+1))^{-1}\{&(\alpha^\dagger(n)\alpha(n-1))^{-1}
(\alpha^\dagger(n+1)\alpha(n))^{-1}\cr &(\alpha^\dagger(n+1)
\sigma_3\alpha(n-1) \}   (\alpha^\dagger(n+2)\alpha(n+1))^{-1}, }\eqno (1.5a)
$$ with
$$
\vec \alpha(n)= (-i\sqrt{\kappa\over2}\Delta\psi^\dagger_n~~,~~\sqrt{2}
(1-{\kappa\Delta\over2} -  {\kappa\Delta^2\over4}\psi_n^\dagger\psi_n
)^{1\over2} )
\eqno (1.5b)
$$
and  at  even $n$ sites different, though  similar expressions hold for
both  $t_n$ and $\vec\alpha(n)$ [8]. The energy spectrum of this model obtained
through the Bethe ansatz is
also rather complicated, and gets simplified only at the continuum limit.
As far as we know, until  now
 not many other proposals were invoked
 to improve this situation [9], particularly to achieve simpler
forms of local conserved quantities at the lattice level. However,
  a completely different approach was
 formulated in ref.[4] through equivalence between NLS and spin models
 using  intertwiners of  quantum spaces.

  Our primary aim here is to consider a  quantum integrable
  lattice model, which at the continuum limit yields the
more general AKNS system [13] and as an allowed reduction the NLS  field model.
The  system considered
fulfills {\it all} the desirable requirements of a discrete quantum system
listed above and most importantly,
exhibits considerably simple expressions for the
related conserved quantities at the lattice level.
Indeed it satisfies the QYBE exactly with the same rational $R$-matrix
  (1.2) allowing also solution via QISM, yields local Hamiltonian
and gives back both the Hamiltonian and
the  Lax operator of the NLS field model at the continuum limit. Moreover,
 it has an  extremely  simple  structure,
which
 induces an almost trivial form for the projector required for the construction
of local Hamiltonians.  Remarkably,
 this projector turns out to be
field-independent and symmetric.
As a further relevant feature, our model also allows a natural vector
generalisation at the lattice level.
 Finally it admits an integrable deformation
involving Tamm-Dancoff type $q$-bosons. On the other hand, for achieving all
these nice properties, one has to pay
some price which is reflected in the non-hermitian
nature of  the physical observables at the lattice level. At the same time
the associated Lax operator lacks the usual $SU(2) (SU(1,1))$ symmetry.

   We should stress here that such Lax operators with
lesser symmetries were found also to be significant in generating a large
 class of quantum integrable  models [11].

\smallskip\medskip

    \noindent {\bf 2. The classical model}\smallskip

     The model under scrutiny may  be given through the Lax operator of the
form
 $$ L(n\mid\zeta) =  \pmatrix {{
  \zeta + \Delta \kappa  \phi(n)
 \psi(n) }
&{- i \phi(n) (\kappa \Delta)^{1 \over 2}
   }
\cr
{i \psi(n) (\kappa \Delta)^{1 \over 2}
   }
&{ 1} \cr}
 ~ . \eqno (2.1) $$
     Its simplified structure compared to (1.3) is explicit, though due to
no-conjugacy relation between between $\phi$ and $\psi$, it is obviously
not hermitian. Note that similar forms of $L$ operators  appear
also in analysing descrete self-trapping systems [14] as well as integrable
systems close to Toda lattice [15].

Recently, the bi-hamiltonian structure of  of the classical
system corresponding to (2.1) has been determined and its
complete integrability has been established [16] rigorously through the
explicit
construction of action-angle variables using the $r$-matrix approach.
Recall that  at the  classical limit the QYBE (1.1) reduces     to
 the classical Yang-Baxter equation
$$ \left \{ L(n\mid \zeta )~\otimes_, ~ L (m\mid \eta )~ \right \} ~=~
[ ~r(\zeta, \eta ) ~,~ L (n\mid\zeta ) \otimes
L (m\mid \eta ) ~] \delta_{nm} ~. $$
For the present model, the quantum $R$-matrix is given by (1.2) and is
related to its classical  counterpart $r$ by:

$${1\over \zeta}{\bf P} R (\zeta) = {\bf I} -i\kappa r(\zeta),
{}~~~~r(\zeta)={\Delta\over \zeta}{\bf P}.$$

    Now to show the transition of the Lax operator  to that of the
     continuum model one should put $
    \zeta=1+i\Delta \lambda$
    and
    $ \psi(n) \rightarrow i \sqrt{ \Delta} \psi(n), \phi(n) \rightarrow -i
\sqrt{ \Delta} \phi(n)$, which introducing
$ ~\psi(n) ~=~ {1\over \Delta } \int_{x_n}^{x_n + \Delta } \psi (x) ~dx ~$,
and a similar expression for $\phi$,
     would yield  from (2.1) : $~
     L(n\mid \zeta) = 1+\Delta {\cal L}(x,\lambda) + O(\Delta^2).$

${\cal L}(x,\lambda)$ is        the Lax operator of
    the corresponding field model, given by
 $$ {\cal L} (x,\lambda) =  \pmatrix {
 { i\lambda
 }
&{ \kappa^{1 \over 2 } \phi
   }
\cr
{ \kappa^{1 \over 2 }  \psi
   }
&{ 0} \cr }.\eqno (2.2)$$
    It may be easily checked that the conserved quantities
    associated with this system are the same as those of the AKNS system
 [13]; moreover, since their Poisson structures coincide, one may conclude
that the two systems are equivalent.
    In fact through a simple gauge transformation
    $$~~~~{\cal L} \rightarrow h {\cal L}h^{-1}  +h_xh^{-1}~~~~,~~~
        h=e^{-i{\lambda\over 2}x}\eqno(2.3),$$
 this
    ${\cal L}$-operator can be changed into the standard Lax operator of
    continuum NLS:
 $$ {\cal L} (x,\lambda) =  \pmatrix {{
  i {\lambda \over 2}
 }
&{ \kappa^{1 \over 2 } \phi
   }
\cr
{\kappa^{1 \over 2 }  \psi
   }
&{-  i {\lambda  \over 2}} \cr     }=
i {\lambda  \over 2} \sigma^3 +
\kappa^{1 \over 2 }\phi \sigma^+ +
\kappa^{1 \over 2 }\psi \sigma^-  ,\eqno (2.4)$$
restoring the unitary symmetry, since as is well known, the
 AKNS system allows the reduction $\phi = \psi^*$.

\smallskip\medskip
\noindent {\bf 3. Quantum model}
\smallskip

Recently,  more general forms of $L$-operator of  discrete quantum integrable
models  corresponding to standard
$R$-matrices    have been proposed [11].
Such class of $L$ operators associated with the rational $R$-matrix (1.2)
and satisfying the QYBE
may be given by the following expression which clearly lacks
the unitary symmetry:
$$ L = \pmatrix {  {K_1 + i  { \lambda \over \kappa } K_2 }  &{K_-}
\cr
  {K_+} & {K_3 + i  { \lambda \over \kappa } K_4 } } , \eqno (3.1) $$
where {\bf K } operators  satisfy the algebra
$$ \eqalign {  [K_+ , K_- ] & =  ( K_1 K_4 - K_2 K_3 ) , ~~[ K_1, K_3 ] = 0
\cr
 [K_1, K_{\pm}] & = {\pm }  K_{\pm} K_2 ,
{}~~ [K_3, K_{\pm}] = {\mp }   K_{\pm} K_4 ,  }  \eqno (3.2) $$
with $ K_2, K_4 $ as central elements. It may be seen that when
$ K_1=-K_3,~~K_2=K_4=1, K_+=(K_-)^\dagger
$,
 (3.2) reduces to the standard $su(2)$ algebra
and one can get back the
known lattice NLS (1.3) through HPT. However
the $L$ operator (3.1) in general gives  the possibility
of generating
other quantum integrable models wich do not exhibit such  symmetry. Quantum
                             Toda chain is one of the main  examples [11].
It is interesting to observe that the quantum version of the
 NLS model (2.1) considered  here,
also falls into this class and can be obtained from (3.1) through
the following realisation:
$$ K_1 = \Delta^2 \kappa \phi \psi + 1,~~
K_2 = - \Delta \kappa ,~~ K_3 = 1,~~K_4 = 0 ,
{}~~  K_+ =  i \Delta \sqrt {\kappa} \psi, ~K_- = -i \Delta \sqrt {\kappa}
\phi,
\eqno (3.3) $$
 where the  operators $\psi, \phi$ obey the canonical commutation relation
$ [~\psi(n)~, \phi(m)~] ~=  {1 \over \Delta}\delta_{nm} ~$.
This
quantum model, represented by the Lax operator
$$
 L(n\mid\lambda)_{NLS} =  \pmatrix {{ 1 -
 i \lambda  \Delta + \Delta^2 \kappa  \phi(n)
 \psi(n) }
&{- i \Delta \kappa^{1 \over 2}\phi(n)
   }
\cr
{i \Delta \kappa^{1 \over 2}\psi(n)
   }
&{ 1} \cr},
\eqno (3.4),$$
as descendant of the integrable  `ancestor' model (3.1), is naturally quantum
integrable  and satisfies the QYBE
with the same $R$ matrix (1.2)
as the
NLS field model.

In exact analogy with the classical case, (3.4) allows transition to
the Lax operator of the AKNS system and through allowed
reduction to that of the continuum quantum NLS model. Indeed,
the gauge transformation (2.3),
being independent of the field operators,
is clearly applicable  to the quantum case as well.

It is known [7] that
the  conserved quantities
$ C_l$ may be obtained from the transfer matrix
$~~~\tau (\lambda) = tr \left( \prod_{N}^{1} L(n\mid\lambda) \right)~~$,
through an expansion at a special point $\nu$, in the form:
$$C_l = {1\over \kappa l!} {\partial^l \over {\partial \lambda^l}} \log \tau
(\lambda)
\mid_{ (\lambda=\nu)}~~~$$.

In what follows we use the method developed in [7,8].
The   locality of the Hamiltonian and other conserved
quantities can be achieved provided that at this special point $\nu$
the operator $L(\lambda)$ be  expressible both as a ``direct" and  an
``inverse" one-dimensional
 projector [7,9]. This in turn implies the vanishing of
 its quantum determinant [6] $det_qL$
at this point, where

   $$\eqalign {det_qL &= tr({\cal P}_-
{}~\left(L(\lambda) \otimes L(\lambda+i\kappa) \right)) \cr &={1\over2}
[(L_{11}\tilde L_{22} +L_{22}\tilde L_{11})-(L_{21}\tilde L_{12}+ L_{12}
\tilde L_{21})]}$$
with   $ {\cal  P}_- = {1\over4}(1-\sum_a \sigma_a\otimes
\sigma_a)$ being the antisymmetriser and $L\equiv L(\lambda)$,~~
$\tilde L \equiv L(\lambda+ i\kappa)$.

We observe that  for Lax operator (3.4)  one  gets $ det_qL= 1-i\lambda\Delta,$
giving a single degeneracy point $\nu_1=-{i\over\Delta}$. The
resulting projector depends on the field operators and one cannot avoid
the  implementation of the  involved procedure
discovered and applied in [6-8] and elaborated in  [9].
Fortunately however, under
an  irrelevant scaling of the Lax operator $L\rightarrow \hat L=({i\over
\lambda\Delta}) L $, which evidently does not affect the QYBE and thus can only
give equivalent lattice models, the quantum determinant becomes
$$ det_q\hat L =-{1\over\Delta^2}\left({1-i\lambda\Delta
\over\lambda  (\lambda +i\kappa)}\right) ={\xi(\xi+\Delta)\over
\Delta^2(1+\kappa\xi)},$$
where $\xi={i\over\lambda}$.
That is, another degeneracy point  $\xi=\nu_2=0$ naturally
appears.

The rescaled operator $\hat L$ takes the form
$$
\hat L(n\mid\xi) =  \pmatrix {{ 1 +
 {N(n)\over  \Delta }\xi
  }
&{- i  \kappa^{1 \over 2}\phi(n)\xi
   }
\cr
{i  \kappa^{1 \over 2}\psi(n)
 \xi  }
&{{1\over\Delta}\xi} \cr},
\eqno (3.4')$$
 with
 $N(k)=  1+\kappa \Delta^2 \phi(k) \psi(k)$.
At the new degeneracy point $\xi=\nu_2=0$,
it becomes remarkably simple as it
turns into a field independent projector:
$$ \hat L(0)=\pmatrix { 1
& 0
\cr
0
&0 \cr}
={\cal P}\eqno(3.5)$$
The above procedure  amounts essentially  to choose the expansion point
at $\lambda =\infty$. We emphasize that the existence of
such an exceptional expansion point where the projector becomes
field-independent is possible only due to the
asymmetry of the present model.
We note incidentally that an analogous property also holds at the classical
level [16].
As a consequence, due to the almost trivial form of $\hat L(0)$ (3.5),
 as we will see now, not only
the required locality condition is satisfied,
but also
the derivation  as well as the expression of
the Hamiltonian and the other conserved quantities become extremely
simple.

For explicit calculations we use now $\hat L$ and expand  around $\xi=0$
assuming periodic boundary conditions,
dropping however the $hat $ sign from all  subsequent expressions.
This gives
$$\tau(0)=tr \left(\prod L(k\mid\xi)\mid_{ (\xi=0)}
\right) = 1,\eqno (3.5a)$$
 $$ \eqalign { {\partial\over\partial\xi} \tau(\xi)_{\mid \xi=0}\equiv
 \tau^\prime (0)&= tr \sum_k \left( L(N\mid\xi)
\cdots   L^\prime(k\mid\xi) \cdots  L(1\mid\xi)\right)_{\mid (\xi=0)}
, \cr  &=  {1\over\Delta} tr   \sum_k \left(   {\cal P} \cdots
 N(k) {\cal P}\cdots
{\cal P} ~\right) \cr
&=  {1\over\Delta}  \sum_k N(k)
. } \eqno (3.5b) $$
 In a similar way one gets
 $$\tau^{''}(0)= 2
 \left( {1\over\Delta^2}  \sum_{i>k} N(i)N(k) + \kappa \sum_k
  \phi(k+1) \psi(k)  \right),\eqno (3.5c)$$
where a factor  $2$ appears  due to the identity
  $$(\cdots L(i)\cdots L'(k)\cdots)'\mid_{\xi=0}
   = (\cdots L'(i)\cdots L(k)\cdots )'\mid
   _{\xi=0}
   $$ valid since
   $L''(\xi)
=0$. Continuing further we get
 $$ \eqalign { \tau^{'''}&(0) = {6\over\Delta}(
{1\over\Delta^2} \sum_{i>j>k} N(i)N(j) N(k)
+ \kappa (\sum_{i,k (i\ne k,\ne k+1)} N(i)
  \phi(k+1) \psi(k)+ \cr &
   \sum_k
  \phi(k+1) \psi(k-1))),}\eqno(3.5d)  $$
  etc.
  Notice that the conserved quantities $C_k$ may be given  through
  the above expressions (3.5) in the following form
   $$C_1={1\over\kappa} (log\tau(\xi))'\mid (\xi=0)=
     {1\over\kappa}     \tau(0)^{-1} \tau'(0)
$$
$$    C_2=
   {1\over2\kappa} (log\tau(\xi))''\mid (\xi=0)= {1\over2\kappa}
   [\tau(0)^{-1} \tau''(0) -(\tau(0)^{-1} \tau'(0))^2] $$

  $$  \eqalign {   C_3&=
  {1\over 3!\kappa } (log\tau(\xi))'''\mid (\xi=0)\cr &=  {1\over6\kappa}
  [ 2 (\tau^{-1}
   \tau'(0))^3+\tau^{-1} \tau'''(0)
   -2 (\tau^{-1} \tau'(0))(\tau^{-1} \tau''(0)) -
   (\tau^{-1}\tau''(0))(\tau^{-1} \tau'(0))], }$$ where $\tau^{-1}$=
   $\tau^{-1}(0)$
 Inserting now the expressions (3.5) one finally gets the required observables
 $$N=C_1=
  {1\over\Delta\kappa}  \sum_k N(k)
 \eqno (3.6a)$$ as the `Number ' operator,
 $$
  P=C_2\equiv\sum_k p_k = \sum_k(
  \phi(k+1) \psi(k) - {1\over 2\kappa\Delta^2} N(k)^2),\eqno(3.6b)$$
  as the `Momentum' operator and
 $$\eqalign { H=C_3&\equiv\sum_k h_k= {1\over\Delta}
  \sum_k (
  \phi(k+1) \psi(k -1) -  (N(k)+N(k+1))
  \phi(k+1) \psi(k)+ \cr &
  (3\kappa\Delta^2)^{-1} N(k)^3 ).}\eqno (3.6c)$$
  as the Hamiltonian of the system.

It may be noted that the above conserved quantities are not
symmetric in $\phi$ and $\psi$, which is a consequence of the
asymmetry of the Lax operator. On the other hand
their locality
  is explicit and
  it is interesting to observe that even though expressions (3.5) given
  through expansion of $\tau(\xi)$ were all nonlocal, in
  the corresponding conserved quantities (3.6) all such nonlocal terms
  get cancelled among themselves leaving only the local ones, as it
occurs also in the classical case.
  We stress again  that the evident simplicity of
expressions (3.6 a,b,c) for the conserved quantities,
 is the most prominent feature of the present model.

  The transition of these conserved quantities
  to those of the
  NLS field model is easily achieved at the continuum limit
  by taking
  $${\bf N}=({1\over\Delta\kappa}  \sum_k (N(k)-1))\mid_{( \Delta \rightarrow
0)}
            = \int dx   \phi(x) \psi(x),\eqno (3.7a) $$
    $${\bf P}=2(\sum_k(p_k+{1\over{2\kappa\Delta^2}}))\mid_{( \Delta
\rightarrow 0)}
            = \int dx  ( \phi_x \psi
            -   \phi \psi_x),\eqno (3.7b) $$
    $$\eqalign {{\bf H}&=-(\sum_k(h_k-{1\over{3\kappa\Delta^2}}))\mid_{( \Delta
\rightarrow 0)}
\cr
    &= \int dx  ( \phi_x \psi_x
            + \kappa ( \phi \psi)^2),}\eqno (3.7c) $$
with the standard assumption of vanishing  boundary condition.
It is worth remarking that the continuous conserved quantities (3.7)
of the AKNS type system are now
symmetric in $\phi$ and $\psi$, which allows therefore the reduction
$\phi = \psi^\dagger$  yielding the   known expressions for the NLS field
model.

The evident closeness between the conserved quantities (3.6) of the
lattice version with those of (3.7) related to the continuum model is
a noticable feature of the present model.
For solving  the eigenvalue problem for the Hamiltonian of the
discrete  model exactly,
we go along the well established steps [8]  of algebraic Bethe ansatz
forming the basic tool of QISM [10].
Defining the monodromy matrix as $ T(\lambda)=\pmatrix {{A
(\lambda)} & {B(\lambda) } \cr
{C(\lambda) } & {D(\lambda) } }=\prod_k^N L(k\mid\lambda) $
we get the expression for transfer matrix as $\tau(\lambda )=tr(T(\lambda))
=A(\lambda)+ D(\lambda)$, which generates the conserved quantities, while
 $B(\lambda), C(\lambda)$     acts as `creation' and `
annihilation' operators, respectively. The
 $n$-particle eigenstates may be defined as $ \mid n > =
\prod_i^n B(\lambda_i) \mid 0> $
with the property of the `vacuum' :
 $$ C(\lambda)\mid 0> =  0,~
A(\lambda)\mid 0>
= a(\lambda)^N \mid 0>,
{}~~D(\lambda)\mid 0> = d(\lambda)^N \mid 0>.    $$

The QYBE for the monodromy matrix is given again by  equation (1.1), with
$L$ operators  replaced by the corresponding $T$ operators. In
elementwise form this equation yields the `commutation' relations
$$\eqalign {&[A(\lambda),A(\mu)]= [D(\lambda),D(\mu)]=[B(\lambda),B(\mu)]
=[C(\lambda),C(\lambda)]=0\cr
&A(\lambda)B(\mu)={1\over c(\mu,\lambda)}B(\mu)A(\lambda) -
{  b(\mu,\lambda)   \over c(\mu,\lambda)}B(\lambda)A(\mu)\cr
&D(\lambda)B(\mu)={1\over c(\lambda,\mu )}B(\mu)D(\lambda) -
{  b(\lambda,\mu )   \over c(\lambda,\mu )}B(\lambda)D(\mu)}\eqno (3.8)$$
with $$b(\lambda,\mu )={i\kappa\over \lambda-\mu-i\kappa},~~
c(\lambda,\mu )={\lambda-\mu\over \lambda-\mu-i\kappa}.~~$$
The eigenvalues of $\tau(\lambda)$ giving the physical observables
may be obtained
by using the  commutation relations (3.8)  between $A,B$ and $D,B$ and
the properties of the `vacuum' stated above.
Skipping out the details we present  only the
main results as follows.
$$ \tau (\lambda) \mid n> =
{\cal E}(\lambda,\{\lambda_j\})\mid n>,\eqno(3.9)$$ where$$
{\cal E}(\lambda,\{\lambda_j\})=
a(\lambda)^N \prod_l^n
{1\over c(\lambda_l,\lambda)}  +  d(\lambda)^N \prod_l^n
{1\over c(\lambda,\lambda_l )}\eqno (3.10)$$
Note that the
form of eigenvalues (3.10) is obtained provided the parameters $ \lambda_i$
satisfy the condition [10]$$
({a(\lambda_j)\over  d(\lambda_j)})^N = \prod_{l\ne j}^n
{c(\lambda_l,\lambda_j)
\over   c(\lambda_j,\lambda_l )  }
,~~i,j=1,\cdots,n.\eqno (3.11) $$

For the present case  one obtains $a(\xi)=(1+{
\xi\over\Delta}), d(\xi)=
{\xi\over\Delta}$. We define the Hamiltonians   by
$$ C_k =
{1\over n!\kappa} {\partial \over {\partial \xi^k}} \log (\tau (\xi) a^{-N}
(\xi))
\mid_{ (\xi=0)},~~~$$
 where $a^{-N}$ is included to remove
irrelevant constant terms and to avoid linear combinations of
conserved quantities. Thus we get from (3.10)
$ C_k =
{1\over n!\kappa} {\partial \over \partial \xi^k} \log
\tilde \tau (\xi)
\mid_{ (\xi=0)},~$ where
$$ \tilde\tau(\xi)=
 \prod_k^n {{1+i( \lambda_k -i\kappa)\xi
} \over {1+i \lambda_k\xi}}
,\eqno (3.12)$$ which finally yields
$$ \eqalign{{\cal N}&=-C_1=n,\cr
{\cal P}&=-i(C_2+{\kappa\over2}C_1) =\sum_{k=1}^n \lambda_k\cr
{\cal E}&=C_3 -{\kappa} C_2 +{\kappa^2\over6}C_1=\sum_{k=1}^n \lambda_k^2
.}\eqno (3.13)$$
We observe that the energy is proportional to $\lambda_k^2$,
the momentum is proportional to $\lambda_k$ and the number of particles
is equal to
the quasiparticle excitation number, as required.
Note again that this result conserning the descrete model
under consideration is much similar to that of the NLS field model [10]
 including the combinations of different conserved quantities to determine
 the momentum and the energy spectrum.
However, contrary to the continuum case, here the values of $\lambda_k$'s are
not arbitrary and should be determined from the equations (3.11), which
for the present  model reads:

 $$
(1-i\lambda_j\Delta)^N = \prod_{l\ne j}^n {\lambda_j-\lambda_k-i\kappa
\over  \lambda_j-\lambda_k +i\kappa  }
,~~i,j=1,\cdots,n\eqno (3.14) $$
We should emphasise that the energy spectrum of this model obtained above
and the related constraints on $\lambda_k$ are
indeed extremely simple.

\medskip \smallskip
\noindent {\bf 4. Vector generalisation of the model}
 \smallskip

 It is interesting to observe that the models violating the $SU(2)$
 type symmetry,
 proposed in [11] can be easily generalised  for the $gl(N)$ case.
 Out of such generalised  system one might then construct
   quantum integrable models, like multi-component Toda chain,
  vector NLS model etc.,
as  realisations through a set of independent  bosonic operators.
 Such  generalised  systems are given by
  the Lax operator
   $$
   L= \sum_l (K_l^+ + {{i \lambda} \over \kappa} K_l^-) e_{ll} +
           \sum_{j \not= l} K_{lj} e_{jl}, \eqno (4.1)$$
 where $(e_{ij})_{kl}= \delta_{ik}\delta_{jl}
 $ are the generators of $gl(N)$. It can be shown that
 the above $L$ operator, associated with the rational $(N^2\times N^2)$
 $R$-matrix : $R(\lambda)= {\bf 1} +i{\lambda \over \kappa}  \Pi
 $, where $\Pi =   \sum_{lk} e_{kl}
 \otimes e_{lk},~$ satisfies the QYBE if the generators ${\bf K} $ yield the
following  algebra:
 $$ \eqalign {
&[K_{mk},K_{kl}]= K_k^-K_{ml}
,~~[K_{kl},K_{lk}]=  K_k^+K_l^- -  K_k^-K_l^+   ,\cr
    & [K_k^+,K_{kl}]=  K_{kl} K_k^- , ~ [K_k^+,K_{lk}]= - K_{lk} K_k^- ,
    \cr &[K_k^{+},K_{lm}]= [K_{kl},K_{km}]=
K_{kl},K_{ml}]= [K_{kl},K_{mn}]=  0, }\eqno (4.2)
   $$ where
 $K_k^- $ commute with all other generators:
  $ [K_k^-,K_{ij}]= 0$
and thus are central elements, while
  $K_l^{\pm} $  form an abelian subalgebra.
 We may notice
 again that in general this is not a $su(N)$ algebra, which
 however is recovered  at some particular symmetric reduction.

 Different  realisations of this algebra would generate through (4.1)
 different quantum integrable models,
  which would  share
 the same rational $R$-matrix but generically would not exhibit
  unitary symmetry.
Consider now a realisation of (4.2) through a set of
 independent  operators with the commutation relations
$ [\psi_l,\phi_k] = \delta_{lk},~~  [\psi_l,\psi_k] = [\phi_l,\phi_k] = 0 ,$
 in the form
 $$  \eqalign {
&K_1^- = -1, ~~ K_1^+ =\sum_j\phi_j\psi_j ,~~ K_i^+ = {\bf 1}_{ii},~
 K_i^- =0,~(i=2,\cdots,N)     \cr
&K_{1j} =\psi_j,~~~   K_{j1} =\phi_j
 ,~ K_{ij} = 0,~~~~~1<(i,j)\le N       }\eqno (4.3)
$$
The corresponding  Lax operator (4.1) will then read
 $$
 L(\lambda) =  \pmatrix {    {
  - {{i \lambda} \over \kappa} +
   (\vec  \phi \vec \psi) }  &
{ \vec {\phi}
   }
\cr
{\vec {\psi}
   }
  &
{ {\bf 1}} \cr }
 .\eqno (4.4)$$
 This Lax operator, which yields a  quantum integrable lattice model,
  gives the vector NLS model [17,18]
   at the continuum limit
 and is a natural
 generalisation of (3.4) to the vector case. The associated
 $R$-matrix  also coincides with that of the field
model [18]. Thus (4.4) is related to the Lax operator  of a new
 exactly integrable
 lattice version of the vector NLS model.
The corresponding classical system has been considered in [16].
 \smallskip\smallskip

 \noindent {\bf 5. A novel quantum integrable Tamm-Dancoff  $q$-bosonic
 model}
 \smallskip  \smallskip
 A number of lattice models involving $q$-oscillators , which are integrable
 at the quantum level have already been discovered [11,19]. Most of these
 models are related to the quantum group structures associated with
 the trigonometric $R$ matrix, which forms a separate class entirely different
 from the NLS model with rational $R$ matrix (1.2).  We present here
 an integrable deformation
 of the discrete NLS model (3.4), which  involves Tamm-Dancoff (TD)
[20] type $q$-boson
 operator, but at the same time is related to a rational $R$-matix.

 It has been shown in [11] that for a 'Symmetry breaking'
 transformation [21]:
$~~ R_{ij}^{kl} (\lambda ) \rightarrow \ \  e^{i\theta (j-k) }
{}~ R_{ij}^{kl} (\lambda ),~~$
of the  original R-matrix (1.2),
where $\theta$ is some constant parameter,
the algebra (3.1) of {\bf K} operators  is  also
 deformed in an interesting
way.
We find a realisation of this deformed algebra through  TD type $q$-bosonic
 operators $b,c,N$:
$$[b,N]=b,~~~ [c,N]=-c,~~~~~~
b c - q c b = q^N ,\eqno (5.1)$$
 where
$b$ and $c$ are not in general hermitian conjugate of each other.
Here we have introduced the parameter $~q=e^{i2\theta}$.
One
may compare the above  TD type  $q$-deformed bosons with
the standard $q$-oscillator algebra [22]:
$~[a,N]=a,~ [a^\dagger,N]=-a^\dagger,~~,
a a^{\dagger} - q a^{\dagger }a = q^{-N} ~$.
 The  algebraic relations    (5.1) yield the
Lax operator of the corresponding   model as
$$
 L^q(\lambda) =  \pmatrix {{ (1+  \kappa  N
- i \lambda ) f(N)}
&{- i \kappa^{1 \over 2}~ c
   }
\cr
{i\kappa^{1 \over 2}~ b
   }
&{f(N)} \cr},
\eqno (5.2),$$ where $f(N)=q^{{1 \over 2}(N-{1 \over 2})}$.

Note that it represents a quantum integrable system, which satisfies
the QYBE with the deformed rational $R$ matrix   $$
  R^q (\lambda) =  \pmatrix {
 { \lambda -i\kappa
  } & { } & { } & { } \cr { } &
{ -i\kappa }& {\lambda q^{-{1\over 2}} }& { } & { }
\cr{ }& { \lambda q^{{1\over 2}}    } &
{ -i\kappa
 }& { } & { }\cr { } & { } & { }
&{ \lambda -i\kappa
  } \cr }
.\eqno (5.3)$$
 Evidently at $~q \rightarrow 1~$
one  recovers
from (5.3) the  Lax operator (3.4) of our discrete NLS and also
gets $R^q \rightarrow R $ as in (1.2).

There is    a simple mapping
from such  TD-deformed operators to the operators of
the original lattice NLS model as $~~b= f(N)\psi ~, c= f(N)\phi~~$
recovering the canonical relation
$[\psi,\phi]=1$; accordingly the
Lax operator (5.2)  is mapped into   (3.4) by:
$$L^q(n) = q^{{1 \over 2}(N(n)-{1 \over 2})} L(n)_{NLS}. $$
Hence, this Tamm-Dancoff type deformed bosonic system
represents a new quantum integrable lattice model which can be
solved
through the algebraic Bethe ansatz using the  results reported
 in sect.3.
 \smallskip\smallskip

 \noindent { \bf 6. Concluding remarks}
 \smallskip
It might be worthwhile to summarize here the main  results
obtained in this paper, and to stress once again its underlying
ideology.
We have presented a quantum integrable lattice model, which in general
corresponds to the AKNS type system. It may also be considered as a
 lattice NLS model, that shares
with the continuum field model the same quantum R-matrix, yields
local and quite simple expressions for the conserved quantities
including the Hamiltonian and allows to
determine their spectrum by the standard quantum inverse scattering
method.
Remarkably, the present  lattice NLS admits a natural vector
generalisation giving   a new exact lattice version
of the vector NLS model, which
is much simpler than all other vector
generalisations available in the literature. Finally, the model
can be easily deformed, giving rise to a  Tamm-Dancoff type q-boson
model, exactly solvable at the quantum level.
The price we had to pay to achieve all the previous results
was the breaking of unitary symmetry, which is however restored
 at the continuum limit. We
 note that the same advantages and the same drawback
characterise the classical version investigated in [16].
 For those who think, as we do, that the price we paid
was an acceptable one, the ideology underlying the construction
of the present model might be a fruitful one.

After the completion of this work  ref. [23,24]  were
brought to our notice. In [23], as a significant contribution,
a most general form of $L$ operator for the lattice NLS was found,
which provides the basis for  classification of all $L$ operators
related to the $R$-matrix (1.2). The physical and mathematical
properties of lattice NLS along with many other models have been
discussed in great detail in [24].
 We note that the model presented here is consistent
with the general form of the $L$ operator of [23], which thus gives
another basis for its validity.
\medskip
{\it Acknowledgements: } One of the authors (AK) acknowledges with thanks
the support of Alexander von Humboldt Foundation. The research reported
in the present paper has been carried out in the framework of the
national research program ``Problemi matematici della Fisica",
supported by the italian Ministery of University and Scientific and
Technological Research
(MURST). The authors are also thankful to Prof. V. Korepin for
essential
constructive remarks and for pointing out the ref. [23,24].

\hfil \break \vfil \eject

\noindent{\bf References }
 \item {[1]} A. Alekseev ,L.D.Faddeev and Volkov CERN TH-5081/91;
 \item {[2]} J.Moser, A.P.Veselov, Comm.Math.Phys. 139 (1991) 217.
\item { } M.Bruschi, O.Ragnisco, P.M.Santini and G.Zhang Tu,
Physica D49 (1991);
        \item {[3]} F.W. Nijhoff, H.W. Capel and V.G. Papageorgiou,
        Phys.Rev. A46 (1992) 2155;
\item { } F.W. Nijhoff,  V.G. Papageorgiou, H.W. Capel, G.R.W.Quispel,
 Inv.Prob. 8 (1992) 597;
 \item { } Yu. B. Suris, Phys.Lett. A145 (1990) 113
\item {[4]} V.O. Tarasov, L.A.Takhtajan and L.D.Faddeev,
Teor. Mat. Fiz. 57 (1983) 163
\item {[5]} M.J.Ablowitz and J.F.Ladik, Stud.Appl.Math. 55 (1976)
 213
\item {[6]} A.G. Izergin and V.E.Korepin,
Sov.Phys.Dokl. 259  (1981) 76
\item {[7]} A.G. Izergin and V.E.Korepin, Nucl. Phys. B 205  (1982)
401.
\item {[8]} N.M.Bogolyubov and V.E.Korepin, Teor.Mat.Fiz. 66 (1986)
455.
\item {[9]} David A.Coker,{\it
Use of projectors for Integrable models of quantum field theory}
\hfil\break Stony Brook preprint, ITP-SB-92-18,June 1992.
\item {[10]} L.D. Faddeev, Sov. Sci. Rev. C1 (1980) 107.
\item {} P.Kulish and E.K.Sklyanin ,
in {\it Integrable Quantum Field Theories, Lecture notes in Physics , }
eds. J.Hietarinta et al. (Springer Verlag, Berlin,1982) Vol. 151, p. 61.
 \item {[11]} A.Kundu and B.Basumallick , Mod. Phys. Lett.  7 (1992)
 61
\item {[12]} V. S. Gerdzikov, M.I. Ivanov, \& P.P. Kulish
J.Math.Phys. 25 (1984) 25
\item {[13]} A.C.Newell, M.J.Ablowitz, D.Kaup, \& H.Segur,
Stud. Appl. Math. 53 (1974) 255.
\item {[14]} V.B Kuznetsov and A.V. Tsiganov, J. Phys. A22  (1989) L73.
\item {} V.Z. Enolskii, M. Salerno,N.A. Kostov \& A.C. Scott,
Phys. Scr.  43 (1991) 229.
\item {[15]} P.L. Cristiansen, M.F. Jorgensen \& V.B. Kuznetsov,
Lett. Math. Phys. 29 (1993) 165.
\item {[16]} I.Merola, O.Ragnisco and G.Zhang Tu, {\it A Novel Hierarchy
of Integrable Lattices},
preprint INFN November 1993, submitted to Inverse Problems.
 \item {[17]} S.V. Manakov, Sov.Phys. JETP,38 (1974) 243.
  \item {[18]} P.P. Kulish, Phys.Scr.20 (1979)  (LOMI preprint P-3-79)
\item {[19]} N.M. Bogoliubov , R.K. Bullough, Phys.Lett. A168 (1992) 264
\item {[20]} I. Tamm, J.Phys. (USSR), 9 (1945) 499
\item {} S. Dancoff, Phys. Rev. ,78 (1950) 382
\item {[21]} M.Wadati, T.Deguchi and Y. Akutsu , Phys. Rep. 180 (1989) 247
\item {[22]}
 A.J.MacFarlane   J. Phys. A22 (1989) 458l
\item {[23]} A.G. Izergin and V.E.Korepin, Lett. Math. Phys. 8 (1984)
259.
 \item {[24]}
V.E.Korepin, A.G.Izergin and N.M.Bogoliubov, {\it Quantum
Inverse scattering Method and  Correlation Functions}, Cambridge Univ.
Press 1993.

\vfil\end